\documentclass[compsoc, conference]{IEEEtran}

\usepackage[english]{babel}
\usepackage[rflt]{floatflt}
\usepackage{graphicx}
\usepackage{url}
\usepackage{multirow}
\usepackage[rflt]{floatflt}
\usepackage{graphicx}
\usepackage{multirow}
\usepackage[table]{xcolor}
\usepackage{supertabular}
\usepackage{spverbatim}
\usepackage{paralist}
\usepackage{multicol}
\usepackage{csquotes}
\usepackage{flushend}
\usepackage{cite}
\usepackage{todonotes}

\usepackage{amsmath}%
\usepackage{MnSymbol}%
\usepackage{wasysym}%
\newcommand{\bl}{\blacksquare}
\newcommand{\sq}{\square}

\usepackage{float}
\usepackage{booktabs}
\usepackage[listings]{tcolorbox}

\usepackage{enumitem}

\usepackage[inline]{trackchanges}

\addeditor{DS}
\addeditor{RA}
\addeditor{SC}
\addeditor{CM}
\addeditor{ML}
\addeditor{AS}

\title{Security, Privacy and Safety Evaluation of Dynamic and Static Fleets of Drones}

\begin{document}
\author{
\IEEEauthorblockN{Raja Naeem Akram\IEEEauthorrefmark{1}, Konstantinos Markantonakis\IEEEauthorrefmark{1}, Keith Mayes\IEEEauthorrefmark{1},\\ Oussama Habachi\IEEEauthorrefmark{2}, Damien Sauveron\IEEEauthorrefmark{2}, Andreas Steyven\IEEEauthorrefmark{3} and Serge Chaumette\IEEEauthorrefmark{4}
}
\IEEEauthorblockA{
\IEEEauthorrefmark{1}Information Security Group Smart Card Centre, Royal Holloway, University of London, Egham, United Kingdom\\
\IEEEauthorrefmark{2}XLIM (UMR CNRS 7252 / Universit\'e de Limoges), Limoges, France\\
\IEEEauthorrefmark{3}School of Computing, Edinburgh Napier University, Edinburgh, United Kingdom\\
\IEEEauthorrefmark{4}LaBRI (UMR CNRS 5800 / Universit\'e de Bordeaux), Talence, France \\
Email: \{r.n.akram,k.markantonakis,keith.mayes\}@rhul.ac.uk\\
\{oussama.habachi,damien.sauveron\}@unilim.fr\\
a.steyven@napier.ac.uk, serge.chaumette@labri.fr}
}

\maketitle

\begin{abstract}
Interconnected everyday objects, either via public or private networks, are gradually becoming reality in modern life -- often referred to as the Internet of Things (IoT) or Cyber-Physical Systems (CPS).  One stand-out example are those systems based on Unmanned Aerial Vehicles (UAVs).  Fleets of such vehicles (drones) are prophesied to assume multiple roles from mundane to high-sensitive applications, such as prompt pizza or shopping deliveries to the home, or to deployment on battlefields for battlefield and combat missions.   Drones, which we refer to as UAVs in this paper, can operate either individually (solo missions) or as part of a fleet (group missions), with and without constant connection with a base station.  The base station acts as the command centre to manage the drones' activities; however, an independent, localised and effective fleet control is necessary, potentially based on swarm intelligence, for several reasons: 1) an increase in the number of drone fleets; 2) fleet size might reach tens of UAVs; 3) making time-critical decisions by such fleets in the wild; 4) potential communication congestion and latency; and 5) in some cases, working in challenging terrains that hinders or mandates limited communication with a control centre, e.g.\ operations spanning long period of times or military usage of fleets in enemy territory.  This self-aware, mission-focused and independent fleet of drones may utilise swarm intelligence for a), air-traffic or flight control management, b) obstacle avoidance, c) self-preservation (while maintaining the mission criteria), d) autonomous collaboration with other fleets in the wild, and e) assuring the security, privacy and safety of physical (drones itself) and virtual (data, software) assets. In this paper, we investigate the challenges faced by fleet of drones and propose a potential course of action on how to overcome them.     
\end{abstract}

\section{Introduction}
\label{sec:Introduction}
Drone technology is developing at a breath-taking pace. From toys for hobbyists, it has now reached a state where both the private and public sectors can rely on them as tools to provide value-added services to users \cite{canis2015unmanned}.  The low cost and mobilisation time of drones are two major drivers behind their adoption.  Checking the structural integrity of a building, for example, is much cheaper with drones than with other aerial platforms, such as helicopters, or with high-rise cranes.  Similarly, drones may be deployed faster than police helicopters -- a potentially attractive feature required by law-enforcement agencies. 

Most likely, however, drones would not operate individually or in isolation; individual drones may form part of a fleet managed by the same organisation, where mission objectives are achievable only if fleet participants cooperate.  Moreover, airspaces may contain drones from other organisations, where individual drones and fleets must negotiate to operate safely.  Another complexity is that the airspace might be governed by a local government authority, which may use an automated management system to handle a dynamic and congested airspace containing large numbers of UAVs.

\subsection{Context}
\label{sec:Context}
There are multiple ways in which UAV fleets can be managed; for example, all decisions might be taken by the fleet management authority (also referred to as control centre).  This requires drones to communicate information to the ground fleet management system, potentially in real-time, and to quickly react to any received instructions.  For certain situations, as discussed previously, drone fleets are likely to act autonomously and should make in-flight decisions without requiring explicit permission from a fleet management control system on the ground. 

Individual drones, with limited computation power, may not be suited for autonomous behaviour if it is based on Artificial Intelligence (AI) techniques -- limited computational and storage resources along with real-time decision requirements imposes just challenges.  
To overcome this limitation, drone fleets may be behave like swarms, where AI algorithms designed for the Swarm Intelligence paradigm can be applied.  For Fleets of Drones (FoD) that use swarm intelligence, we refer to them as `Swarm of Drones (SoD)'; the rest are referred to as `FoD'.  An important consideration in swarm intelligence is the nature of the swarm and relationship between swarm-objects, particularly the concept of individual objects entering and leaving the swarm.  This is crucial for a SoD that might either be static, dynamic or hybrid fleets.  In a static fleet, all drones remain together for the whole duration of the mission.  In dynamic fleets, individual drones may enter and leave the fleet as necessary or required by their mission. Lastly, hybrid fleets have a partial fleet behaving statically with the rest as a dynamic fleet.

\subsection{Challenges and Problem Statement}
\label{sec:ChallengesAndProblemStatement}

For a SoD, there are core operations that are necessary for successful mission completion.  These include flight control, flight routing, obstacle avoidance, regulation conformation (regarding the airspace usage), and self-protection (including safety-critical operations) with respect to physical integrity and cybersecurity.  All of these operations are highly time-sensitive. Collectively, we refer to these time-sensitive operations as `Mission-Critical Operations' (MCO). The listed operations are loosely related to three main challenges discussed below.

\subsubsection{Swarm Intelligence}
\label{sec:SwarmIntelligence}

Swarm intelligence takes inspiration from the collective behaviour of natural systems, such as swarms of insects; these systems are inherently decentralised and often have the ability to self-organise.
The ability of a swarm of insects to perform certain tasks emerges out of the interaction of simple and quasi-identical agents, which act asynchronously due to the lack of a central control \cite{Beni2005}.
Algorithms based on swarm intelligence principles, like ant colony optimisation, bee-inspired algorithms and particle-swarm optimisation, are used in optimisation problems that are static or change over time.


In distributed robotics systems, a swarm or fleet of autonomous agents may operate in remote locations with little or no control by a human operator.  Swarm robotics uses large numbers of autonomous and situated robots with local sensing and communication capabilities \cite{Brambilla2013}.


A swarm of robots offers certain advantages over the use of a single one.
Due to the large number of robots, the workload can be distributed across the swarm, and multiple tasks can be worked on simultaneously \cite{Barca2013}.
It further offers distributed sensing capabilities and an increased robustness to failure, by eliminating the single point of failure, as demonstrated in the Swarmanoid project \cite{Dorigo2013}.

The swarm's behaviour is often optimised using evolutionary algorithms; for instance, researchers have successfully evolved a swarm's ability to adapt to unknown environments \cite{Urzelai2001,Bredeche2009}, its resilience to failure \cite{Millard2014a}, and the planning and following of formation patterns \cite{Saska2014,Dorigo2013}.


\subsubsection{Security, Safety and Privacy}
\label{sec:SecrutiyAndPrivacy}

Any flying asset -- drones in this case -- can be a potential target to harm its current state or to access the data it contains, whether it is part of a group or individual.  These two elements raises the challenges of how to implement the security of the asset so its current state cannot be compromised \cite{javaid2012cyber,gupta2016survey}, and how the data can be protected in a manner so it does not violate any privacy requirements \cite{Gregory2013}.  Similar issues regarding intentional hacking and signal jamming, accountability of security issues, management/enforcement of airspace restrictions, and concerns over privacy and intrusiveness were detailed in a report to the US Congress \cite{Elias2012}.  Furthermore, not just national governments are concerned with the security, safety and privacy of drones flying in cities -- the US Congress passed legislation covering UAVs development and integration in civilian airspace (PUBLIC LAW 112–95—FEB. 14, 2012) -- but also general public \cite{Chang:2017,Lidynia2017,Cavoukian2012}.  This is alongside a number of companies trialling the deployment of drones as part of their on-demand services, e.g. Amazon for Prime Delivery \cite{amazondrone2016}.

Individual drones and FoD, therefore, require strong assurances in terms of security, safety and privacy.  There are multiple options in which the assurances and countermeasures can be built for FoD.  One option is to opt for a set of static policies defined before the FoD commences its mission.  This is useful if the FoD operates in an static environment that has fixed and predictable behaviour; creating fixed policies for MCOs is an obvious choice here.   In reality, however, drones operating in the wild\footnote{Wild: Environment that is not under the control of the drone operators/owners.} has the potential to present scenarios that were not considered previously by drone owners and operators.   For this reason, in this paper, we forward the proposal of designing SoDs based on swarm intelligence.  All of the MCOs would have a deep foundation in swarm intelligence and have the potential to collaboratively learn, evolve and decide the best course of action when operating in the wild, without depending on a ground fleet management systems.

\subsubsection{Performance and Energy Consumption}
\label{sec:PerformanceEncergyConsumption}

Drones are resource-constrained pieces of equipment; individual aircraft are heavily impaired by limited processing capabilities and severe battery or fuel constraints.  It is widely acknowledged that drone power consumption, whether it uses a thermal or electrical engine, is a major issue\cite{serge:DBLP:conf/iros/AbdillaRB15}.  These energy constraints influences all the parameters of the drone system, as well as the mission itself.  The impact of MCOs on energy management is important: on one hand, the MCOs impact the energy consumption -- for instance, because of encryption algorithms\cite{serge:article} -- but the power management must also take MCOs into account, e.g.\ reserving enough energy to ensure prompt responses to critical external stimuli that require quick (and energy demanding) route changes.  Using a swarm is beneficial to ensure that the overall energy burden is shared, e.g. sharing processing loads (see below), to maintain continuous flight and mission succession.

Regarding performance in terms of computing power, even though the technology is evolving very quickly, the processors embedded on small drones (which usually constitute swarms) are not the most efficient.  It should be noted that this lack of computational power also comes from the power management issue (see above): the more efficient a processor, the more power it requires. Therefore, to achieve a significant level of performance, load sharing (in addition to highly tuned algorithms) is required. Load can be shared between the drones of the swarm themselves or between the drones and some external system (a bigger drone or even a ground system). For instance in terms of image processing, mosaicking is often used \cite{serge:doi:10.1117/1.JRS.10.016030}. It consists of taking several photos of a given area and then assembling them to build a global picture that can thereafter be processed depending on the situation at hand. Such a process can be shared among the individual drones of the swarm, dispatching the load all over them \cite{serge:unknown,serge:chaumette:hal-01391871}.

 The two challenges described in this section crucially depend on swarm intelligence, which, in turn, impacts power consumption and computational capabilities. All challenges should thus be considered in a holistic approach.

\subsection{Contributions}
\label{sec:Contributions}
The paper contributes in three main aspects to further the discussion on the management of autonomous and independent FoDs that are:

\begin{enumerate}
\item A rationale supporting application of drone fleets and potential impact of building a SoD. 
\item A conceptual architecture for the SoD, its different variants based on the swarm (enrolment) structures and collaboration models.
\item Finally, charting the open issues that impact SoD in general but specifically the security, safety and privacy of SoD. 
\end{enumerate}

\section{Related Work}
\label{sec:RelatedWork}
In this section, we discuss the related literature from three aspects. 

\subsection{Swarm Robotics - Experiencing, Learning and Adaptation}
\label{sec:SwarmRobotics_ExperiencingLearningAndAdaptation}

Swarm intelligence is not a new concept for FoD. Existing literature \cite{wei2013agent,purta2013multi,Madey2014,7303086} has already explored different uses of swarm intelligence in the context of FoD, especially in the case of internal swarm communication and route planning. However, in related literature, it is difficult to find a case where swarm intelligence is proposed for all operations ranging from flight control to cybersecurity -- as is the case of this paper.

The swarm intelligence paradigm has been used to optimise and control single UAVs:
In \cite{Ross2013}, single vehicle autonomous path planning by learning from small number of examples.

In \cite{Wang2016}, three-dimensional path planning for a single drone using a bat inspired algorithm to determine suitable points in space and applying B-spline curves to improve smoothness of the path.

In \cite{Couceiro2013}, authors introduced and validated a decentralised architecture for search and rescue missions in ground based robot groups of different sizes.
Considered limited communication ability with a command centre and employs distributed communication.

In \cite{Pugh2006}, distributed unsupervised learning in a swarm of robots using a particle-swarm optimisation algorithm.
Accounting for limited communication capabilities amongst members of the swarm.
Requires \emph{a priori} knowledge of the terrain.


In \cite{Saska2014}, the authors achieved area coverage for surveillance in a FoD using visual relative localisation for keeping formation autonomously.

In \cite{Vasarhelyi2014}, authors explored the use of swarm intelligence paradigm to control formation flight and stabilisation through the use of GPS and locally shared information.

In \cite{Madey2014}, authors have investigated the use of a communication middleware and a rule based system to command and control an otherwise autonomous FoD.


\begin{figure*}[htbp]
\hfill
\centering\raisebox{-0.6\height}{\centering\includegraphics[width=.49\linewidth]{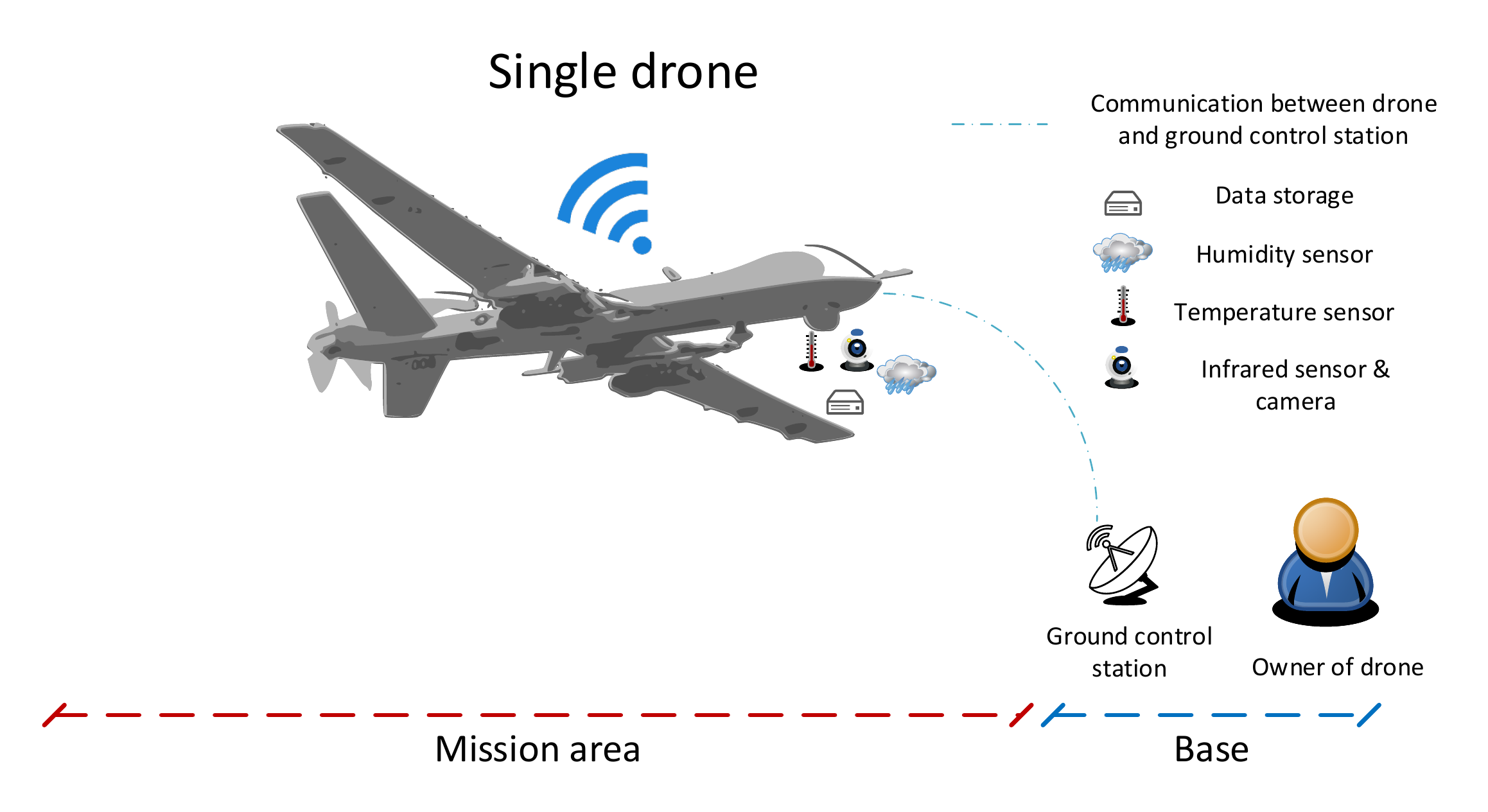}}
\hfill
\centering\raisebox{-0.5\height}{\centering\includegraphics[width=.49\linewidth]{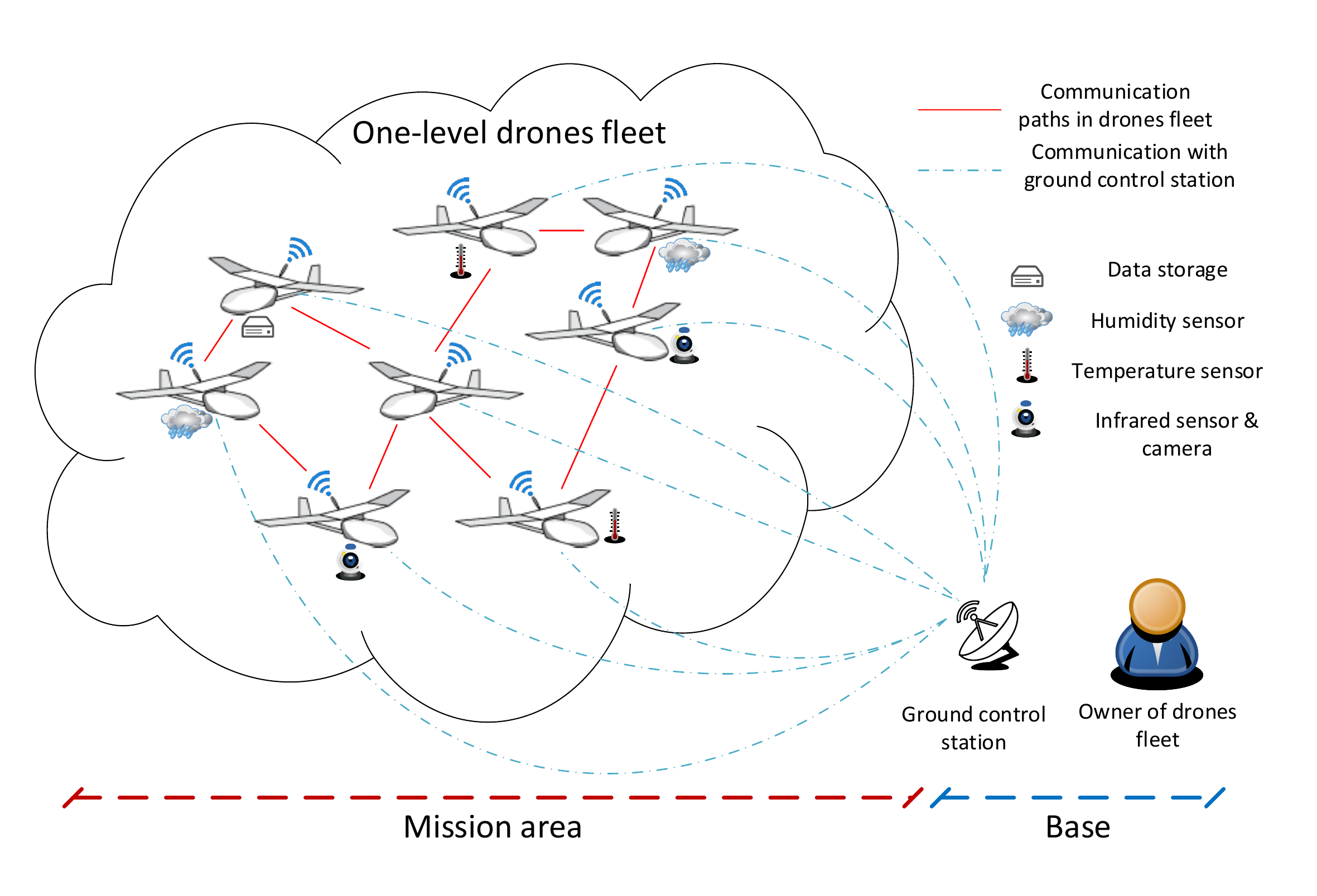}}
\hfill		
	\caption{Single drone versus a one-level drones fleet}
	\label{fig:single-vs-fleet}
\end{figure*}

\subsection{Security, Safety and Privacy of Fleet of Drones}
\label{sec:SecurityAndPrivacyOfFleetOfDrones}
Since the emergence of drones, different papers have proposed solutions to secure them and their communications: either a), an alone drone communicating with a GCS (Ground Control Station) or with other devices, or b), communication inside a FoD.
As stated in~\cite{AkramTrustCom2016d}, the attack vectors to consider are either the capture of drone to make physical or logical attacks, or attacks through its communication capabilities -- all of which might be conducted by a highly sophisticated adversary.

\subsubsection{Individual Drones}
For individual drones controlled by a GCS, the authors of~\cite{Steinmann2016ICNS} proposed a protocol to secure communication along with ensuring that illegitimate accesses to sensing data is not easily available to an attacker. While this proposal provides the deniability property to help to deal with privacy issues (i.e. a GCS is not able to prove to other parties from which drone the message was received), it is expensive to implement in terms of computation and thus energy consumption.

In a similar context, a more efficient proposal relying only on lightweight primitives was done by the authors of~\cite{Blazy2017ICNS} to establish a secure channel protocol when GCS is in the communication range of drone. Their proposal ensures confidentiality and privacy-protection of collected data; if a drone is captured, data cannot be accessible by an adversary.

In~\cite{Won2015AsiaCCS}, the authors proposed a secure communication protocol between drones and smart objects based on efficient Certificateless Signcryption Tag Key Encapsulation Mechanism using ECC, which addresses issue of drone capture.

In~\cite{mtita2017dasc}, the authors used drones to perform efficient inventory and search operations over some RFID tagged assets using lightweight secure and privacy-preserving serverless protocols defined in~\cite{mtita2016efficient}, while guaranteeing the privacy of the tags and the secret when the drone is captured (i.e. compromised). 

In~\cite{Akram-ICNS2017a}, the authors proposed a secure and trusted channel protocol to enable communication of a drone and sensors of Aircraft Wireless Networks (AWNs) to retrieve collected data and ensuring their confidentiality.

\subsubsection{Drones in Fleet}
In the HAMSTER (HeAlthy, Mobility and Security based data communication archiTEctuRe) solution for unmanned vehicles~\cite{Pigatto2016}, the authors presented a security framework and cryptographic schemes without discussing specifically of secure channel protocols and issues of captured drones. 

In~\cite{Maxa2016DASC}, the authors proposed a secure reactive routing protocol, called SUAP, for a fleet of drones. The proposal is efficient to detect and prevent routing, e.g. wormhole and blackhole attacks, but it does not consider an adversary with a high attack potential nor the issue of captured drones.

In~\cite{AkramTrustCom2016d}, the authors proposed to address an adversary with a high attack potential by adding a secure element to each drone of the fleets.  Based on the built architecture, in~\cite{Akram2017wistp}, they proposed a secure and trusted channel protocol to establish a secure channel between the communicating drones and to provide security assurance that each drone is in the secure and trusted state.

\subsection{Performance and Energy Consumption}
\label{sec:PerformanceAndEnergyConsumptionStateOfTheArt}
Energy management is addressed at two different levels: in terms of refuelling (fuel or batteries) capabilities and in terms of in-flight/mission power consumption optimization. Regarding refuelling,
research and experiments are being done to provide mechanisms to reload/refuel during the flight. Standard avionic procedures are being used/adapted but more original approaches are also explored, like the usage of solar panels, laser power beaming or ad hoc hosts \cite{serge:abebe2017drone} for instance. Regarding in-flight power consumption optimization, the drone (internal) supervision algorithms  (the algorithms that control the sensors, the IMU, the autopilot, \emph{etc.}) are studied, as well as the algorithms used to achieve/implement the missions. For instance, flight path management, sense \& avoid, \emph{etc.} can highly impact power consumption.

Computational load sharing is addressed mainly by the computing community, rather than the electronics community. Still, it should be noted that strong relationships are required with the situation management people so as to determine the important information that are really required for the decision process\cite{serge:cummings2007operator}.  This is to avoid the burden and thus the processing load of processing potentially useless data.

It should also be noted that managing and organizing a swarm induces an overhead in terms of computational power and  energy consumption. Indeed, this requires additional communication and management of swarm-related data (location of drone, proximity, RSSI, \emph{etc.}). Even though power consumption has been addressed in some work (see for instance \cite{serge:DBLP:conf/syscon/BrustAT16}) it is still an issue to consider. Moreover, security and safety related features also impact power consumption and computational load, and it is necessary to take these into account.

\section{Fleet of Drones - Why?}
\label{sec:FleetOfDrones}

\begin{figure*}[htbp]    
\centering
\includegraphics[width=0.90\linewidth]{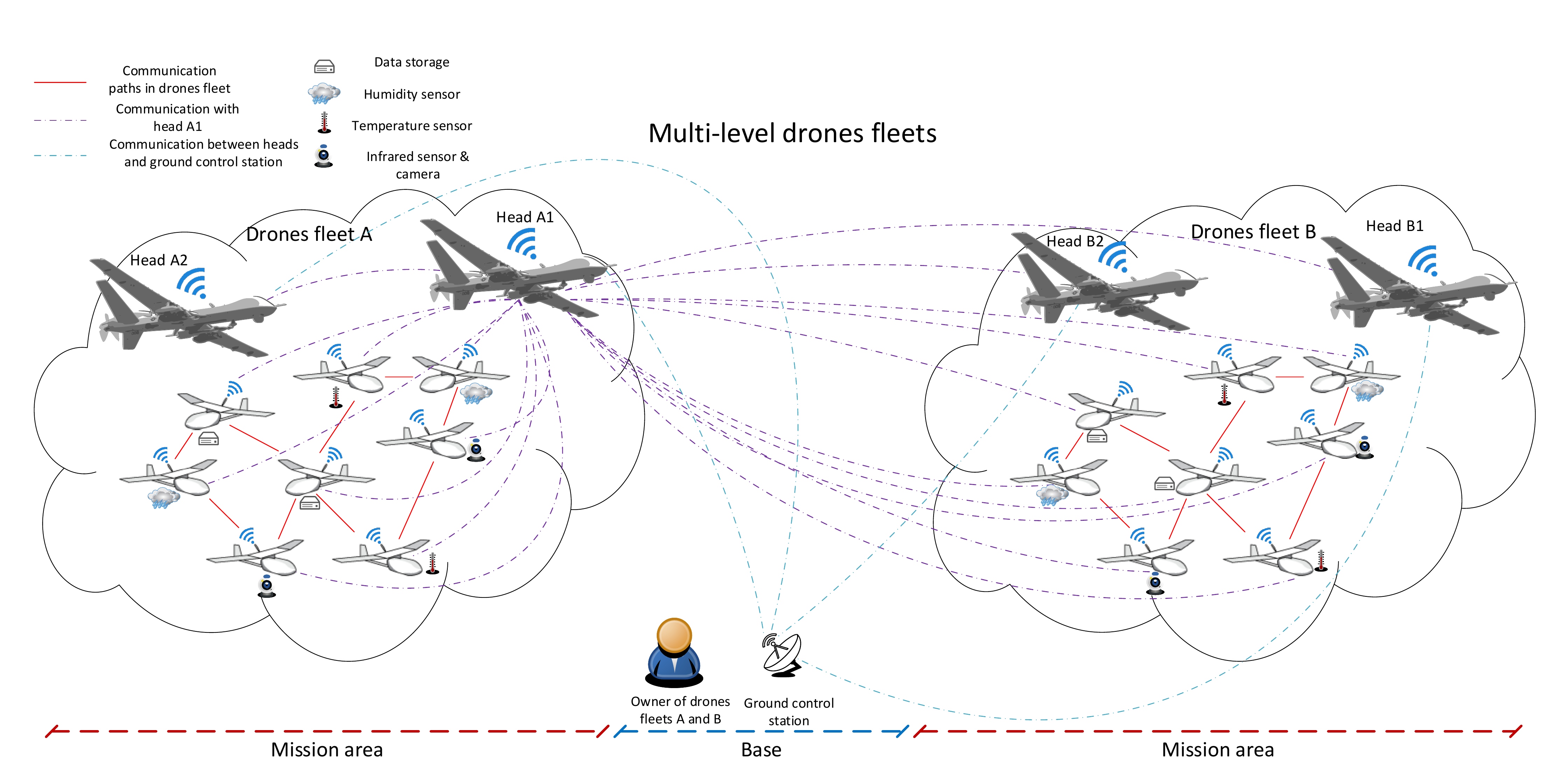}
	\caption{A multi-level drones fleets}
	\label{fig:multi-level-fleets}    
\end{figure*}

Why should FoD can be preferred to single, powerful drones? As illustrated Fig~\ref{fig:single-vs-fleet}, one of the first interest of the fleet is to be composed of several smaller drones that can be equipped with different sensors or other equipments providing redundancy that can help to tolerate a certain degree of failure. In addition, a multitude of drones can cover a larger geographic area than a single one.  This can be performed in a smarter way, since only required drones with useful capabilities to support the mission can be sent to specific areas while other drones perform other tasks. Nevertheless, a single drone would have to attend each location. FoD can also take advantage of the network they form altogether to continue to communicate with the GCS when there are obstacles on the path between some drones and GCS by simply relying messages to neighbour drones which are able to establish communication with GCS. 
Small drones are also interesting because they are more stealthy, less noisy than a big drone that might be of interest for both military and civilian applications.
Last, but not least, small drones can be less expensive than large drone due to their mass-production, and they might be safer for use in civilian applications since their weight is smaller, which is safer in the event of a crash.

However, large drones and smaller versions should not be opposed since they can be used in a complementary way; for example, in a multi-level fleet in which they can serve as relay (they can also be seen as cluster heads) for the smaller drones to enable communication with a GCS or with other FoDs. In Figure~\ref{fig:multi-level-fleets}, only a 1-level fleet is depicted.

\begin{figure}[htbp]
\includegraphics[width=\linewidth]{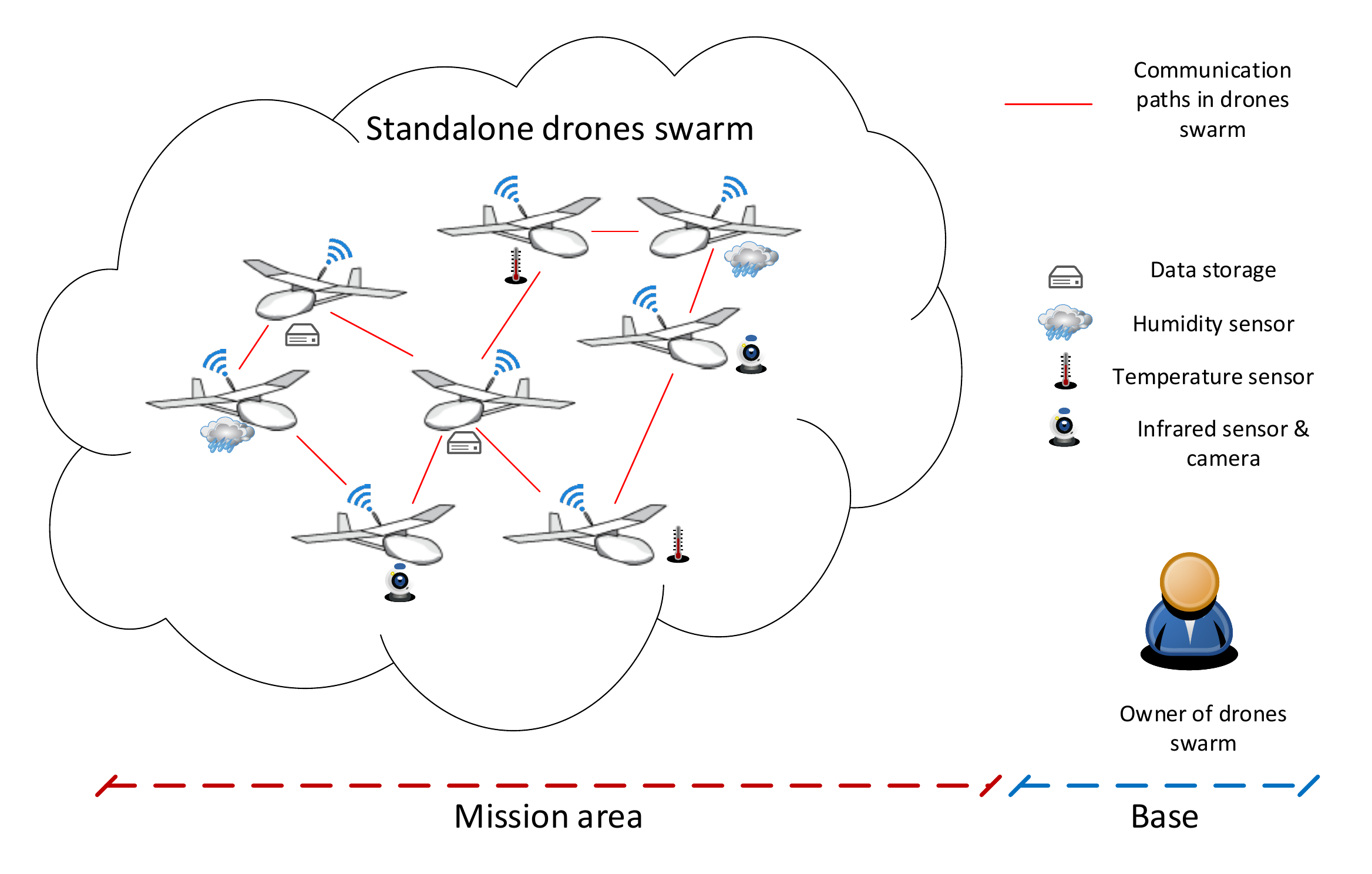}
	\caption{A standalone swarm of drones}
	\label{fig:drone-swarm}    
\end{figure}

It is worth noting that FoDs presented before always received command from the GCS. Of course, they can have a certain degree of autonomy but standalone swarm of drones, illustrated in Figure~\ref{fig:drone-swarm} acting like swarm of animals/insects can be regarded as highly desirable for researchers and operators.  Indeed, once the mission is given, a SoD no longer requires to be driven by a GCS, hence making it autonomous and more stealthy. 

\subsection{Fleet of Drones - Commercial, Civilian and Military Need}
\label{sec:FleetOfDrones}
Depending on the end-users, different architectures of FoDs can be envisioned.

In a commercial context, FoDs can be shared by several stakeholders to decrease the cost of each having its own fleets. For instance, it can be imagined FoDs spread over countries to achieve the entitled missions on-demand of stakeholders that would have to pay according their uses. The missions can be of different types, delivery (assets buy in on-line shops, pizza, drugs, etc.), monitoring of fields or herds for agriculture sector, surveillance of buildings. The main requirement is a good cost effectiveness for FoDs, i.e. the fleets can be able to be shared between several stakeholders in a fair usage and in a dynamic way to enable drones of a fleet to join drones of other fleets to achieve objective of the missions. For such an applicative context, a certain degree of autonomy can be useful but essentially a WAN infrastructure of communication will be used. e.g. LTE network, to control the FoDs.

In a civilian context, authorities would require safe and reliable FoDs for rescue operations (fire detection in forest, searching missing people in sliding snow or after an earthquake, etc.), for smart city scenarios (detection of traffic offences, hunt of criminals, etc.), to monitor major infrastructures (nuclear plants, pipelines of oil and gas, power lines, water reserves, airports, railways, etc.) and borders.  In this context, most applications can use WAN communication infrastructure to control the FoD, in addition to the drone-to-drone communication inside the FoD, to exchange data and potentially control commands to achieve the mission. One of the scenarios where communication infrastructures might no longer exist is after a disaster (earthquake, tsunami, hurricane, huge terrorist attacks) that destroyed them. Authorities can also use FoDs to fight against malevolent drone flying in forbidden areas by capturing it.

In a military context, requirements are stealth, protection of data regarding the missions (flight plans) and the data collected (positions of interests), adaptability to the adverse condition when deployed in the field. For stealth, SoDs are the best since they do not use the wide range communication which minimises potential detection by the adversaries. The SoD is adaptive to adverse conditions, to failure or destruction of some drones, to fulfil the missions. In any case, FoD or SoD of drones are more resilient than single drone for most of applications in this context. 

It is worth noting that the security and privacy protection are requirements shared by all contexts.

\subsection{Use Cases}
\label{sec:UseCases}

This section present three use cases for FoDs and SoDs.

\subsubsection{Rescue Operation in Remote Areas}
\label{sec:RescueOperationInRemoteAreas}
FoDs and SoDs can be used in several rescue operations in remote areas where an event -- earthquakes, for instance -- has made access difficult or dangerous for emergency services. They can also be used to set up a network of communication dedicated to emergency staff for data exchange or to recover public networks (like 3G/4G) to enable victims and other people to communicate with their families or urgency staff.
FoD can also be used in case of sliding snow to cover a wide area than human persons to search victims. In addition, such small drones can fly closer to the ground than big drone.  This can more efficient to detect signs of life; for instance, to look for people at sea after a plane crash.  FoDs can cover a larger area than conventional means (helicopters, planes and ships).  For such scenario, autonomy in energy is an issue to deal with.

\subsubsection{Facility Surveillance and Fault Detection}
\label{sec:FacilitySurveillanceAndFaultDetection}
As mentioned previously, FoDs can be used for surveillance of wide area to detect abnormal event.
For instance, they can detect fire in natural parks or forests by covering wide area equipped with multiple sensors (e.g. thermal) and they can also fight it with embedded dedicated payload to extinguish the flames before they grow.
For buildings and any large infrastructure requiring a high security, FoDs can provide an additional security level by providing a third dimension in defence against intrusion or degradation. Indeed the drones help to detect intruders based on embedded sensors and camera. However they can also provide a fourth dimension by being able to fight promptly against intruders, potentially at the price of their own destruction. For instance, recently at the time of the writing, a few hundred dollars drone land on the deck of HMS Queen Elizabeth -- without anyone raising the alarm.  It can be imagined that one or several drones of a FoD protecting this ship would have tried to capture and/or to destroy the intruder's drone by sacrificing them if required. 
For border surveillance, FoDs can help cost efficiency by avoiding continuous human patrols or wall construction.

\subsubsection{Data Collection}
\label{sec:DataCollection}
FoDs can also help to collect data from wireless sensors nodes (WSN) which are not always connected to a sink having a permanent internet connection (for instance, it can be a WSN requiring stealthy since it operates on an adversary field or individual sensors positioned on the ground but that do not form a network to save their energy). In such applications, the drones of the fleets are somehow mobile sinks to retrieve collected data. This kind of uses of FoDs can exist in smart city scenarios where to avoid crowding of radio frequency spectrum, sensors nodes disseminated in the city may only emit with a very low power that requires the recipient is very close to collect the data; which such small drones can do.

For data collection tasks, drones of a fleet can be used for inventories of RFID tagged assets if they are equipped with RFID readers, but also of livestock in wide areas using cameras.

Finally, a basic and common scenario of data collection with drones that can be extended to FoDs is ground imaging capture for different purposes, such as for military (to find point of interests: position of enemies for instance) and agriculture (to view area requiring watering, or those requiring treatment against a disease).

\section{Swarm of Drones - Technology Perspective}
\label{sec:FleetOfDrones}
In this section, we discuss a conceptual architecture that can deployed for SoD.

\subsection{Generic Architecture for SoD}
\label{sec:PotentialArchitecturesOfFleets}

\begin{figure}[htbp]
\centering
\includegraphics[width=\linewidth]{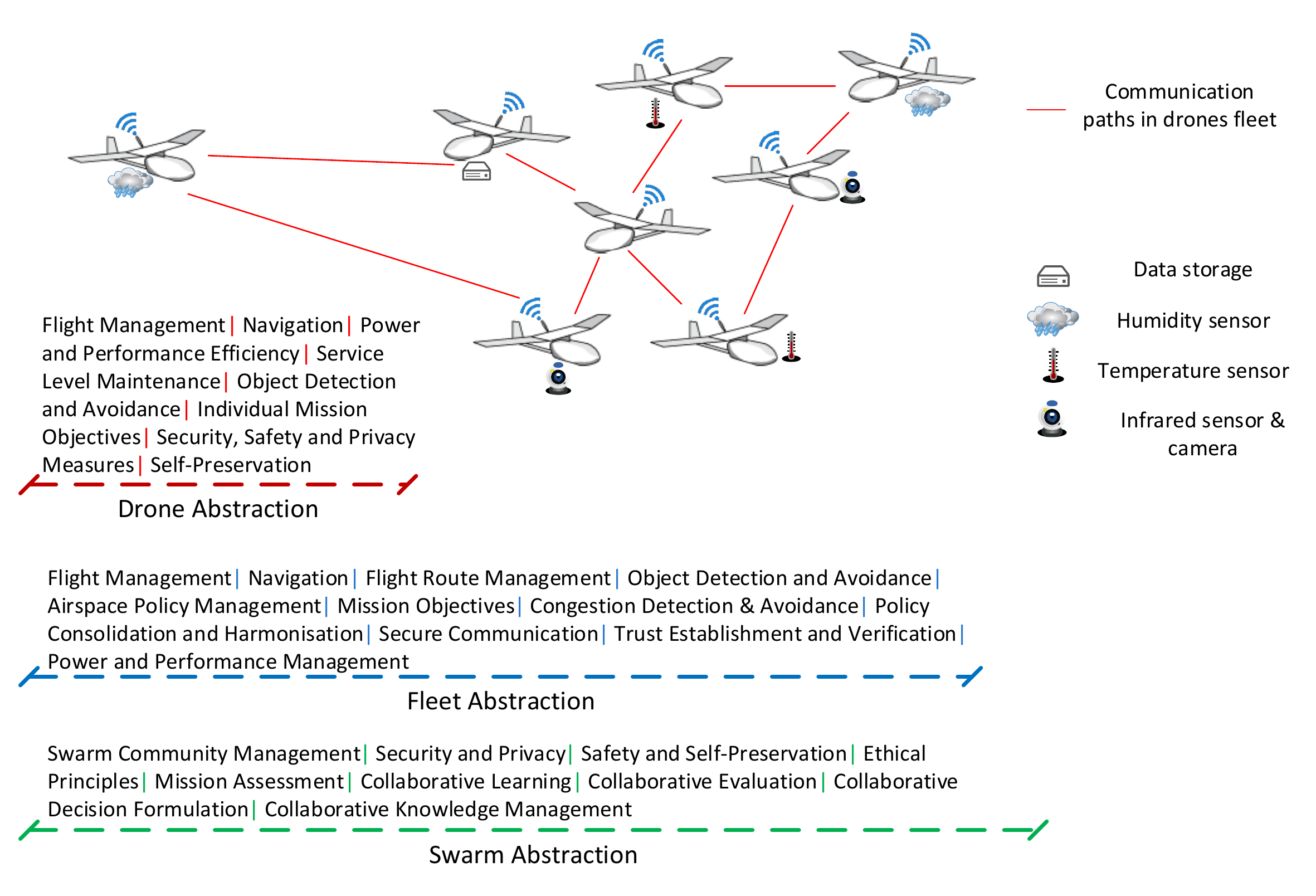}
	\caption{Conceptual architecture of SoD}
	\label{fig:ConceptionArchitectureofSoD}    
\end{figure}

The conceptual architecture shows the set of operations in two different contexts: 1), how they are stacked in a single drone, for example, operations are that specific (or individual) to a drone and how it is related to other operations on the drone, and 2), how different operations are actually a collaborative options in which the swarm decides rather than individual drones. Figure~\ref{fig:ConceptionArchitectureofSoD} shows the conceptual architecture. The architecture is divided into three layers with some duplication and each layer; the rationale of these is discussed in subsequent sections. 

\subsubsection{Drone Abstraction}
\label{sec:ApplicationAbstraction}
This abstraction layer is focused on a single drone operations, which preserves the drone as an individual entity and includes:

\begin{enumerate}[label=D\arabic*]
\item Flight Management: This operation ensures that the drone remains in the air as required by the mission. The flight management features can be semi-static or partly dependent on \ref{itm:F-FM} and can only be modified in unique circumstance if required by \ref{itm:S-CDF}. \label{itm:D-FM}
\item Navigation: This involves airborne movements within the respective SoD, in relation to the external environment, flight route, airspace authority's injunction and \ref{itm:D-ODA}. Each drone has a static set of rules for its navigation activity; however, these can be superseded by the collaborative decision making process of the SoD. \label{itm:D-Nav}
\item Power and Performance Efficiency: This operation continually monitors in the drones power and performance matrix -- and at set intervals notifies the swarm. In case the power and performance efficiency has severely degraded, the conflicting self-preservation \ref{itm:D-SP} and SoD preferences (\ref{itm:S-SSP} and \ref{itm:S-CDF}) kicks in and based on the severity and mission requirements the drone can either take the route of disengaging from the mission or altruism. \label{itm:D-FE}
\item Service Level Maintenance: Similar to the \ref{itm:D-FE}, this operation looks into the entirety of the services a drone is offering as part of the SoD. Any delays or difficulty to fulfil its obligations, required by the SoD, it will raise the notification for the SoD (\ref{itm:S-MA} and \ref{itm:S-CL}), so adequate mitigation can be applied via \ref{itm:F-PPM}, \ref{itm:S-CE} and \ref{itm:S-CDF}. \label{itm:D-SLM}
\item Object Detection and Avoidance: This operation has two aspects, first to detect a obstacle approaching the flight path and second the avoid collision. These two actions are dual nature, due to time criticality of this operation, first option is that the decision might be taken by a drone individually, but even in this case it would notify the remaining swarm. The second option is that obstacle is not detected by the drone itself but by other swarm member and it takes adequate measure to avoid it pre-emptively. \label{itm:D-ODA}
\item Individual Mission Objectives: At the time of swarm construction (section \ref{sec:FleetConstruction}), the SoD owner would upload the objectives of the mission. These objectives would be specific to two levels, individual drones and the fleet (\ref{itm:F-MO}). This information would detail the criticality of mission, responsibility of individual drones and the fleet as a whole. This information would be used by \ref{itm:D-SP}, \ref{itm:S-SSP} and \ref{itm:S-MA}. \label{itm:D-IMO}
\item Security, Safety and Privacy Measures: This operation monitors the individual drone level security, safety and privacy features. The baseline rule set can be pre-defined either solely based on SoD owners design or an autonomously evolved formulation (from a baseline) based on the collaborate knowledge of all SoD flights (carried out by the respective SoD or other SoDs in the past). If a situation appears that an individual drone has not encountered before, then it can raise that to the SoD to take a collaborative decision (\ref{itm:S-CDF}). \label{itm:D-SSPM}
\item Self-Preservation: Depending upon the criticality of the mission, role of a drone and analysis from \ref{itm:S-MA}, a drone might opt for selfish attitude to preserve its operational integrity over the requirements of SoD or opt the altruism approach. In the later approach, individual drone might came to the decision to sacrifice its operational integrity for the success of the mission or in corner condition to upload the ethical principles (\ref{itm:S-EP}). \label{itm:D-SP}
\end{enumerate}

\subsubsection{Fleet Abstraction}
\label{sec:Task/MissionAbstraction}
This abstraction layer bridges between the decisions taken by individual drones on their own and the course of action that is stipulated as part of the mission brief from the SoD owner, with feed in from the swarm abstraction layer in case an unexpected situation is encountered in the wild.

\begin{enumerate}[label=F\arabic*]
\item Flight Management: The operation that is configured to manage the flight operations of the SoD as per pre-defined mission brief. The flight management operation is mission focused and pragmatic as depending on the situation it would either opt for pre-defined plan or \ref{itm:S-CDF}.\label{itm:F-FM}
\item Airspace Policy Management: This function of the SoD remains in constant communication with airspace controller and other drones operating in the same space to comply with the regulations stipulated in the respective region. When making decisions, whether by individual drone or by the SoD as a whole, it consults this function and abide by the airspace regulations.\label{itm:F-APM}
\item Navigation: This functions manages the airborne movements of the fleet as a whole based on the \ref{itm:F-APM} and follows the feeds of \ref{itm:F-FRM}. \label{itm:F-NV}
\item Flight Route Management: The route planner for the whole of the fleet is triggered by either \ref{itm:F-APM} and \ref{itm:F-CDA} -- but remains inside the airspace regulation (\ref{itm:F-APM}).\label{itm:F-FRM}
\item Object Detection and Avoidance: Logically, at the fleet abstraction layer this is part of the \ref{itm:F-FRM} but discussed separately. It depends upon individual drones detection, notification to the population in the SoD and potential avoidance strategy formulation (fleet level) -- based on the analysis results of \ref{itm:S-CDF}.\label{itm:F-ODA}
\item Mission Objectives: Manages the fleet wide mission objectives that are configured at the point of swarm construction (discussed later). This function assists multiple functions during the normal flight, however, in unique situation the swarm abstraction layer takes over to make adequate modification for successful completion or abortion of the mission.\label{itm:F-MO}
\item Congestion Detection and Avoidance: Based on the drone sensors and/or external  feed like airspace traffic broadcasts, this function would identify potential congestion on the selected route of the mission. Based on this detection, it can notify the flight management (\ref{itm:F-FM}) to take adequate actions.  \label{itm:F-CDA}
\item Secure Communication: Manages the set-up and maintenance of secure communication channels between drones in the SoD and with external entities. \label{itm:F-SC}
\item Trust Establishment and Verification: Depending upon the type of SoD (section \ref{sec:TypesOfFleets}) this functionality would establishes the trust relationship between drones in the SoD and with external entities.  \label{itm:F-TEV}
\item Policy Consolidation and Harmonisation: Depending upon the type of the SoD (section \ref{sec:TypesOfFleets}) either all drones would abide by a single policy (covering airspace regulations, ethical principles and swarm participation guidelines) or they have different, sometime conflicting policies. When a drone enrols into a SoD this operation verifies whether the enrolled drone is compatible with the baseline policy of the SoD or not. \label{itm:F-PCH}
\item Power and Performance Management: Computation and power are two scarce resources for the SoD. This operation continuously monitors individual drones state and performs load-balancing to achieve maximum contributions from individual member of the SoD.\label{itm:F-PPM}
\end{enumerate}

\begin{figure*}[htbp]
\centering
\centering\includegraphics[width=0.90\linewidth]{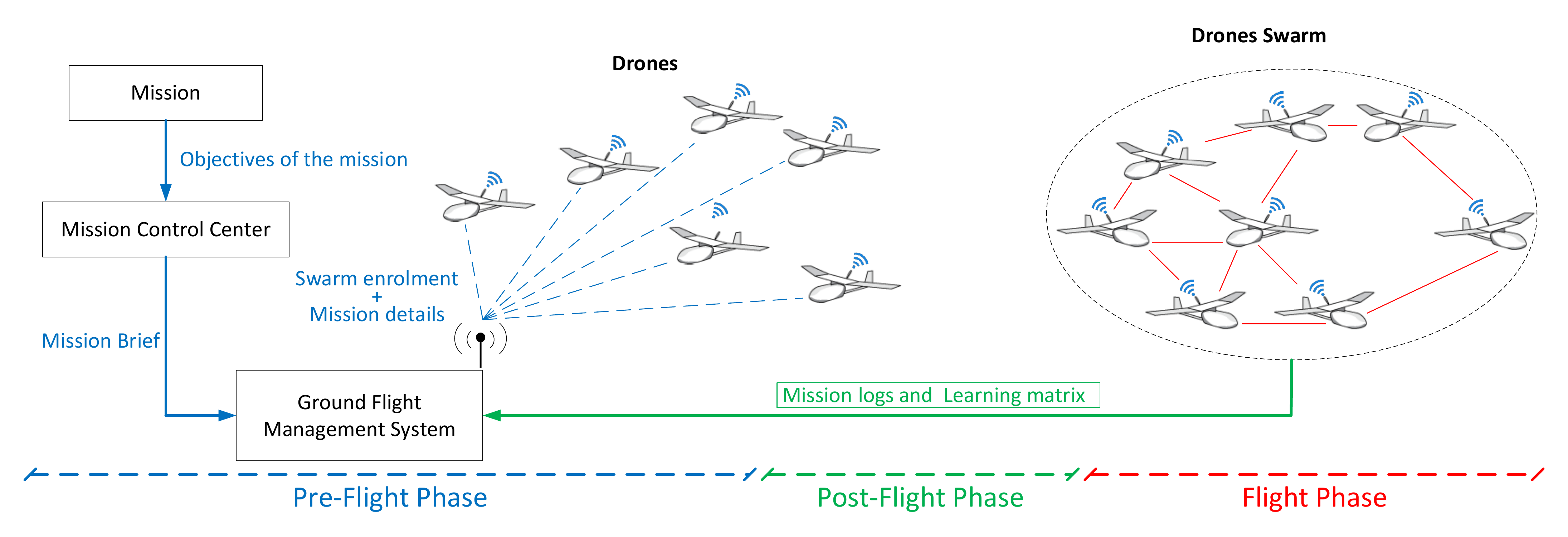}
	
	\caption{Swarm of drones construction process -- pre- and post-mission activities.}
	\label{fig:SoDConstruction}
\end{figure*}

\subsubsection{Swarm Abstraction}
\label{sec:SwarmAbstraction}
This abstraction layer is the foundation of the SoD proposal. The services in this layer, similar to the other abstractions layers, are continuously running on individual drones. This layer has a baseline knowledge: a collection of knowledge that is accumulation of all the SoD flights managed the SoD owner/operator. Therefore, learning, evaluation and decision formulation performed during a single mission then becomes part of the collaborative knowledge to improve all future missions.  

\begin{enumerate}[label=S\arabic*]
\item Swarm Community Management: This service manages the drones participating in the SoD, their contributions and also detects any potential free-riders. \label{itm:S-SCM}
\item Security and Privacy: Deals with the unique situations encountered by the SoD that are specific to the security and privacy-preservation.\label{itm:S-SP}
\item Safety and Self-Preservation: Similarly to the \ref{itm:S-SP}, this service deals with safety and self-preservation of individual drones and SoD as a whole.  \label{itm:S-SSP}
\item Ethical Principles: Set of ethical principles that are set by the drones owner. The \ref{itm:S-CDF} will take these principles into account when making decision.\label{itm:S-EP}
\item Mission Assessment: Swarm health feeds collected by the \ref{itm:S-SCM} will be used by the mission assessment service to perform the prediction of failure and success of the whole mission. This prediction would be useful to make challenging decision whether to continue the mission or abort it. Furthermore, this analysis would be part of the decision to take the mission abortion or an altruistic\footnote{Altruistic Decision: Sacrificing few of the members of the SoD to achieve overall mission objectives} decision. \label{itm:S-MA}
\item Collaborative Learning: The core module that continuously learns from different feeds that is being shared in the SoD. One point needs to be emphasised that the learning process is also collaborative -- as each of the drones might not have the resources to perform this entirely by itself.    \label{itm:S-CL}
\item Collaborative Evaluation: Based on the learning, the SoD would evaluate a situation collaborative to see whether a precedent exists in the collaborative knowledge, if not the collective make a decision autonomously (\ref{itm:S-CDF}). \label{itm:S-CE}
\item Collaborative Decision Formulation: The decision formulation service that requires collaboration from the SoD participants to reach a decision either based on the existing knowledge or take a trail-error strategy. The decision taken and its success would be recorded along with the situation parameters -- post-mission evaluation and inclusion to the collaborative knowledge management (further discussed in section \ref{sec:FleetConstruction}). \label{itm:S-CDF}
\item Collaborative Knowledge Management: One of the objective of the SoD is to accumulate the knowledge from every mission to a single collaborate knowledge based that can then be part of every subsequent missions -- exploring as many as possible permutations of scenarios that SoD can encounter in the field.\label{itm:S-CKM}
\end{enumerate}

\subsection{Fleet Construction}
\label{sec:FleetConstruction}
Based on the conceptual model, the first step is the formation of the SoD at the pre-mission stage and deformation at the post-mission stage. Therefore the process of fleet construction consists of two parts: pre-mission and post-mission, which are discussed in this section. 

At the pre-mission stage, the fleet construction process begins with the formulation of a mission -- with a set of objectives. The mission control unit generates a mission brief that includes mission objectives, airspace regulations, ethical principles, security and privacy policies, organisation commitments, baseline configuration (for first mission), and collaborative knowledge. The mission brief is then communicated to the ground flight management system (GFMS). This system would select the drones from the inventory that would participate in the mission. This selection process is based on the mission requirements, drone availability and organisation preferences. Once the set of drones are selected, the GFMS would then upload the mission brief to the selected drones. Once the brief is uploaded, the drones would establish secure communication channel among themselves in the SoD. Once all drones are connected and GFMS has given the permission to commence the mission, the SoD would initiate. 

After the completion of the mission, upon return of the SoD participants to the base, the GFMS will connect with each drone to download the mission logs, learning/evaluation matrix and potential material that can contribute to the collaborative knowledge. The GFMS communicates this information to the mission control centre that would analyses the mission debriefing information and improves the collaborative knowledge. Figure~\ref{fig:SoDConstruction} shows the fleet construction process with both pre- and post-mission activities. 

\subsection{Types of Swarm of Drones}
\label{sec:TypesOfFleets}
In this section, we discuss three types of SoDs that can be potentially deployed depending upon the target environment and situation. 

\begin{figure*}[htbp]
\hfill
\centering\raisebox{-0.4\height}{\centering\includegraphics[width=.49\linewidth]{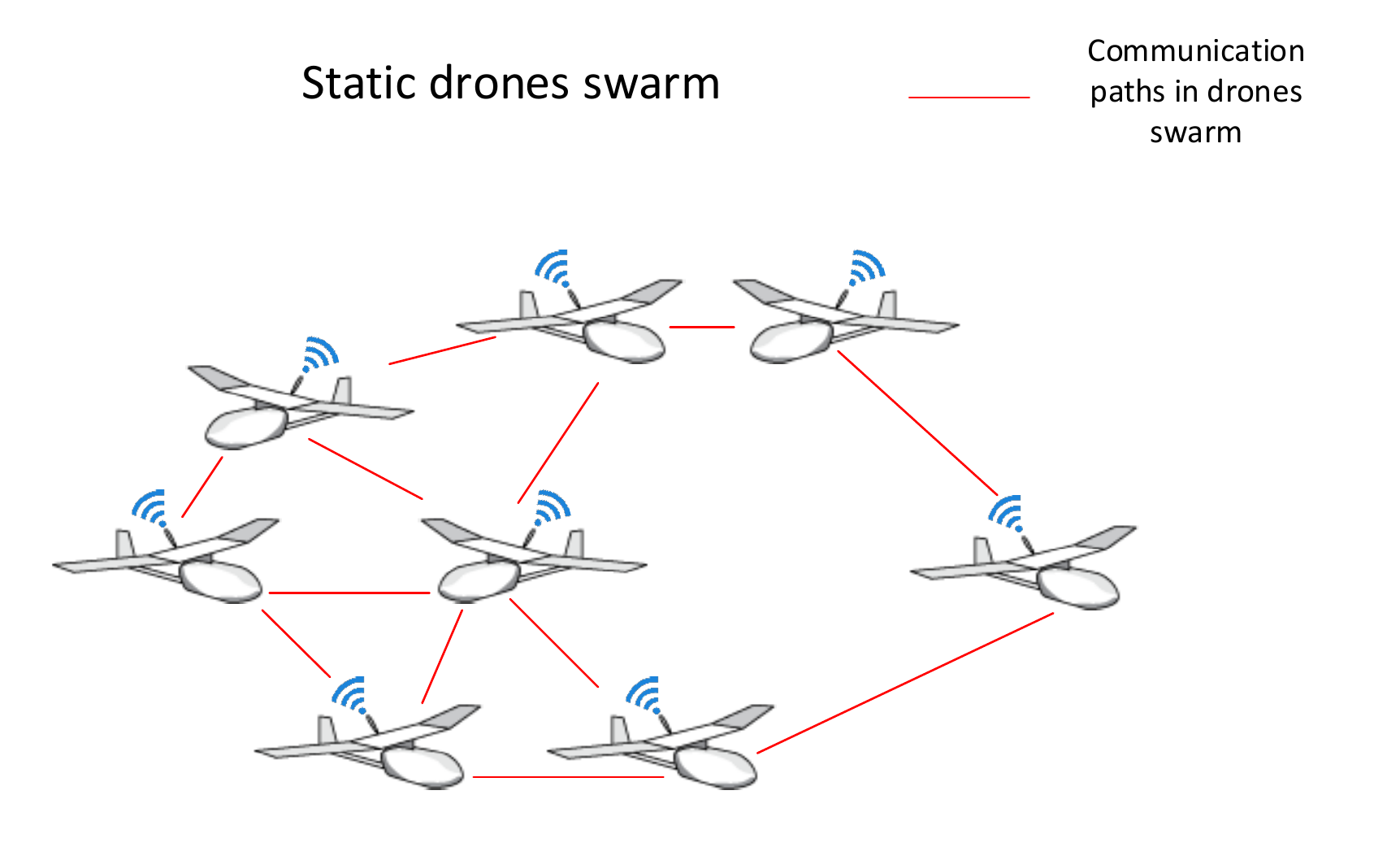}}
\hfill
\centering\raisebox{-0.5\height}{\centering\includegraphics[width=.49\linewidth]{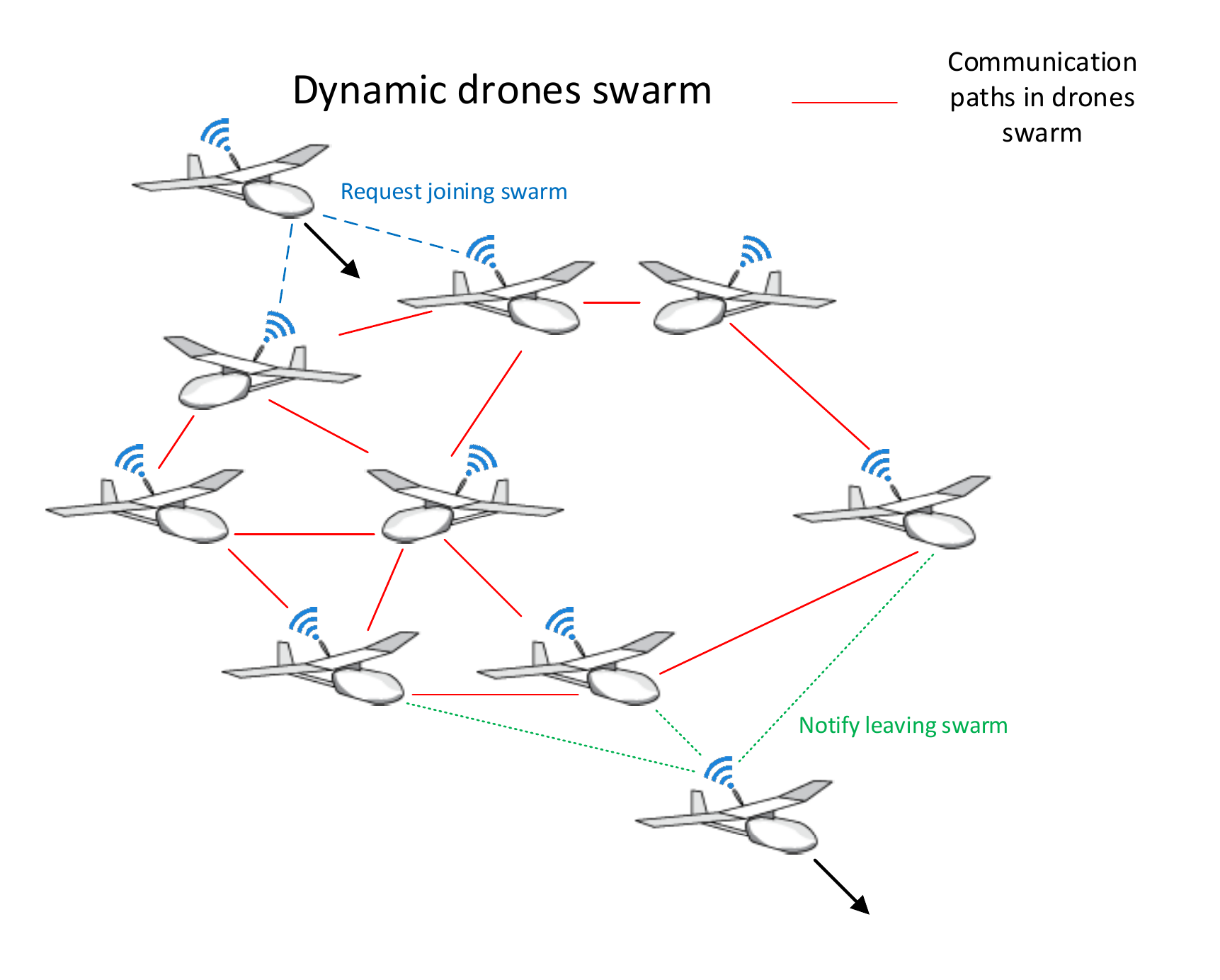}}
\hfill		
	\caption{Static swarm-of-drones versus dynamic swarm-of-drones}
	\label{fig:static-vs-dynamic}
\end{figure*}

\subsubsection{Static SoD}
\label{sec:Static}
The most basic type of the SoD is the static SoD. In this formation, the members of the swarm are pre-selected at the pre-mission stage. During the flight, no new members can enrol into the swarm as the collective is locked at the point of mission commencement. The secure communication, mutual-trust and collaboration is setup by the GFMS of the SoD owner. Any drone that is whether belong to the respective SoD owner or not would be treated as an external entity to the SoD during the flight. Figure~\ref{fig:static-vs-dynamic} shows the static SoD in comparison with the dynamic SoD discussed in the next section. 

\begin{figure}[htbp]
\centering
\includegraphics[width=\linewidth]{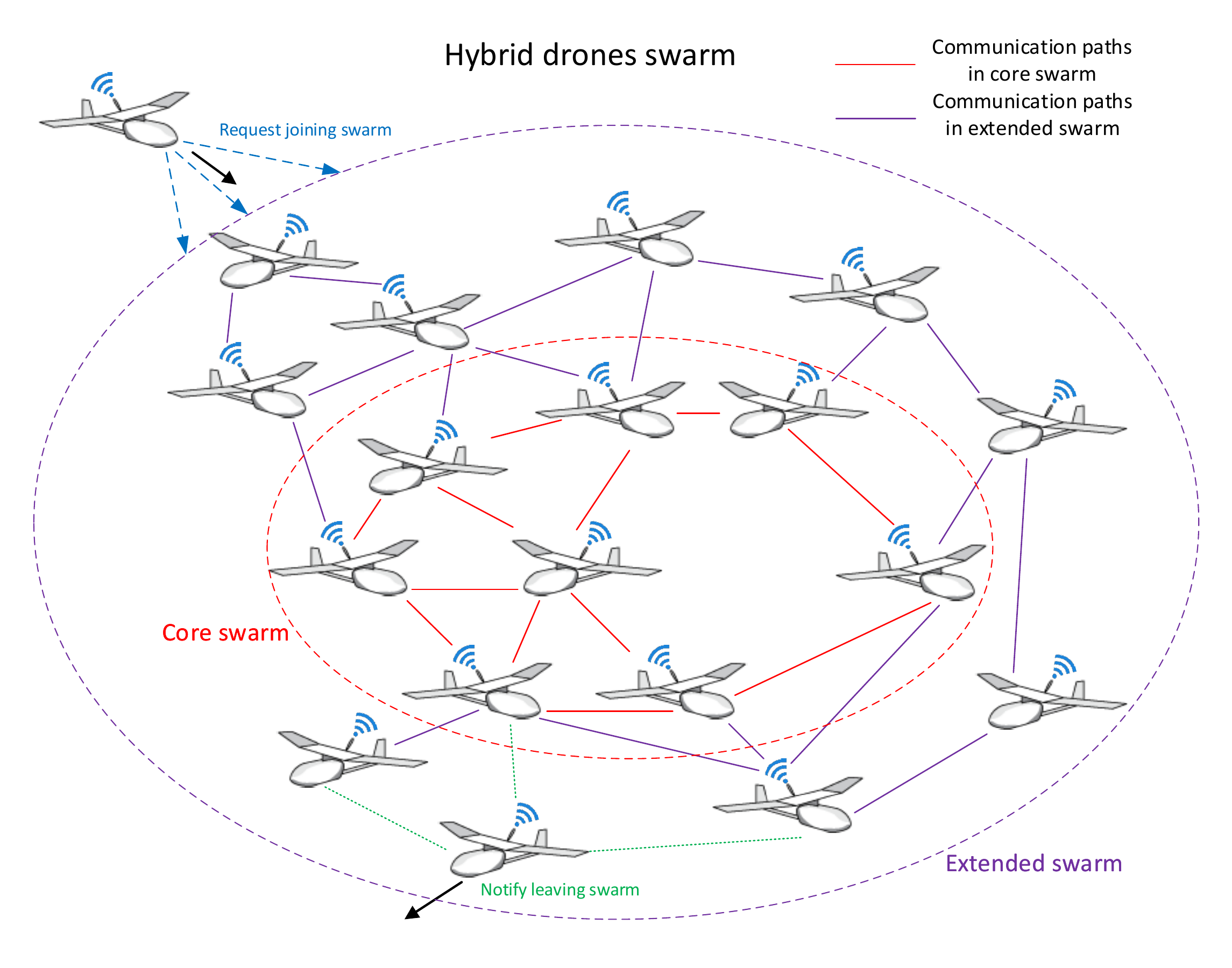}
	\caption{A hybrid swarm of drones}
	\label{fig:hybrid-drone-swarm}    
\end{figure}

\subsubsection{Dynamic SoD}
\label{sec:Dynamic}
In contrast to a static SoD, a dynamic SoD is open to the inclusion of new members along with existing members leaving the swarm at any point of time: pre-mission and/or during the mission. Such a SoD can either be a closed dynamic SoD that only allows enrolment of new drones from the same organisation, or an open dynamic SoD that allows enrolment of drones from any third-party organisations. Whichever the case, the challenges of secure communication, mutual trust and collaboration are unique in comparison to static SoD.

\subsubsection{Hybrid SoD}
\label{sec:Hybrid}

This variant of SoD combines both the static and dynamic SoDs together into a single collaborative unit. At the core of this SoD is a static SoD that behaves like one in all of its operations.  This static SoD, however, is open to allowing other drones to join the swarm, thus creating an extended swarm that behaves like a dynamic SoD. One thing to note is that the core swarm takes high priority when making any collaborative learning, evaluation and decisions. The extended swarm can be viewed as drones that join the static SoD and would provide a service to the core swarm in return for some fair exchange. Members of the extended swarm can leave the collective at any stage. Figure~\ref{fig:hybrid-drone-swarm} shows the hybrid SoD construction.

\subsection{SoD Collaboration Models}
\label{sec:TypesOfSwarmDronesCollaborationModels}
In this section we discuss three variants of collaboration models for the SoD. 

\begin{enumerate}
\item Centralised: In this collaboration model, there is a powerful, master drone in the SoD that collets all the feeds from individual drones and assists the swarm in computing and agreeing on decisions. 
\item Decentralised: In this collaboration model, there is no single master drone, but instead a small subset of powerful drones that collect the feeds from their neighbouring drones and then these powerful drones perform the collaborative learning, evaluation and decision. 
\item Distributed (Peer-Oriented): In this model, each and every participant of the SoD have more or less equal role in all collaborative learning, evaluation and decision making process. It can be noted that the activity load is distributed among the population based on their individual capabilities, current performance and power resources and the criticality of their unique features to overall mission. 
\end{enumerate}

\section{Open Challenges of Swarm of Drones}
\label{sec:OpenChallenges/Problems}
In this section, we selected a very short list of open problems of the SoD. They are categorised into two categories: 1), Security, Privacy and Trust related, and 2), Performance and Energy Consumption related. The implication and importance of these open issues are listed in Table~\ref{tab:ImpOpenChallengeCombination}, represented by number of $\bl$; the higher the number, the more crucial this open issue is to the success of the SoD. 

\subsection{Security, Privacy and Trust Related}
\label{sec:SecurityPrivacyAndTrustRelated}

\begin{enumerate}[label=SP\arabic*]
\item Swarm Authentication, Attestation and Secure Communication: SoDs have to negotiate with external entities that include the airspace controllers, and other UAVs (includes SoDs). Beside this, for dynamic and hybrid types of SoDs, the inclusion of new drones and de-listing of ones that leave the swarm is another challenge. This open issue does not impact the static SoD as much it does the other two types.  
\item Fair Exchange Services Architecture for Swarm of Drones: When drones participate in a group supporting swarm intelligence-based mechanisms to create a SoD, individual drones should have this synergy worthwhile. This should not tax them to the extend that solo mission is comparatively less costly and have negative impact on performance and energy of the drone than if it does not participates in the SoD. Fair exchange becomes way more relevant in the case of dynamic and hybrid SoD as during the flight swarms would potentially be changing and to identify the benefits of joining a swarm has to be clear and verifiable -- use fair-exchange mechanisms. 
\item Collaborated Cybersecurity Deterrence Mechanism: This open issue concerns with how swarm intelligence can be deployed to provide a wide range of countermeasures -- protecting individual drone and the whole of SoD. This is still an open issue and potentially the most crucial element of the SoD proposal.  
\item Detecting the Mole and Free-Riders in the Swarm: This open issues relates to the SP2, however, it focuses on detecting free-riders in the SoD. A free-rider is a drone in the SoD that does not contribute its fair share and becomes a burden on the rest of the drones in the SoD. 

\end{enumerate}

\subsection{Performance and Energy Consumption}
\label{sec:PerformanceAndEnergyConsumptionRelatedChallenges}

\begin{enumerate}[label=PE\arabic*]
\item Balancing the Cybersecurity with Performance and Energy Consumption: we have seen in section \ref{sec:PerformanceEncergyConsumption} that the computational power and the energy consumption of a drone is highly impacted by
several factors. Among these, it is clear that MCOs (including critical event response capabilities), algorithms, data management (ciphering  for instance) are of utmost importance. All these aspects must thus be mitigated with the cybersecurity issues in a holistic approach.

\item Graceful Degradation -- Altruism versus Selfish-Survival: Graceful Degradation is of course part of the intrinsic management system of each drone.
Still, when combined as a swarm, this should not longer be considered only at the level of one single drone but at the level of the swarm as a whole. Indeed, depending on the mission, it must be decided if it is more important for each drone to save its own energy/computational power (selfish approach) or if cooperating (and thus sharing energy consumption/computational load) is more appropriate to ensure the success of the mission (altruist approach).
\end{enumerate}

\begin{table}[tp]
	\centering
	\caption{Importance of Open Challenge/Problem to Combination of SoD Types and Collaboration Models.}
	\label{tab:ImpOpenChallengeCombination}
	\resizebox{0.968\columnwidth}{!}{%
		\begin{tabular}{@{}ccccccc@{}}
			\toprule
			& \textbf{SP1} & \textbf{SP2} & \textbf{SP3} & \textbf{SP4} & \textbf{PE1} & \textbf{PE2} \\ \midrule
			\multicolumn{7}{c}{\cellcolor[HTML]{EFEFEF}\textbf{Static SoD}} \\
			\multicolumn{1}{l}{\textbf{Centralised}} & $\bl\bl\sq\sq\sq$ & $\bl\sq\sq\sq\sq$ & $\bl\bl\bl\bl\sq$ & $\bl\sq\sq\sq\sq$& $\bl\bl\bl\sq\sq$& $\bl\bl\bl\bl\sq$\\
			\multicolumn{1}{l}{\textbf{Decentralised}} & $\bl\bl\bl\sq\sq$ &$\bl\sq\sq\sq\sq$ & $\bl\bl\bl\bl\bl$ & $\bl\sq\sq\sq\sq$ & $\bl\bl\bl\bl\sq$ & $\bl\bl\bl\bl\bl$ \\
            \multicolumn{1}{l}{\textbf{Distributed}} & $\bl\bl\bl\sq\sq$ & $\bl\sq\sq\sq\sq$ & $\bl\bl\bl\bl\bl$& $\bl\sq\sq\sq\sq$ & $\bl\bl\bl\bl\sq$ & $\bl\bl\bl\bl\bl$ \\
			\multicolumn{7}{c}{\cellcolor[HTML]{EFEFEF}\textbf{Dynamic SoD}} \\
			\multicolumn{1}{l}{\textbf{Centralised}} & $\bl\bl\bl\bl\sq$ & $\bl\bl\sq\sq\sq$  & $\bl\bl\bl\bl\bl$ & $\bl\bl\bl\bl\sq$ & $\bl\bl\bl\bl\sq$ & $\bl\bl\bl\bl\sq$ \\
			\multicolumn{1}{l}{\textbf{Decentralised}} & $\bl\bl\bl\bl\sq$ & $\bl\bl\bl\sq\sq$  & $\bl\bl\bl\bl\bl$ & $\bl\bl\bl\bl\bl$ & $\bl\bl\bl\bl\bl$ & $\bl\bl\bl\bl\bl$ \\
            \multicolumn{1}{l}{\textbf{Distributed}} & $\bl\bl\bl\bl\sq$ & $\bl\bl\bl\sq\sq$ & $\bl\bl\bl\bl\bl$ & $\bl\bl\bl\bl\bl$ & $\bl\bl\bl\bl\bl$ & $\bl\bl\bl\bl\bl$ \\
			\multicolumn{7}{c}{\cellcolor[HTML]{EFEFEF}\textbf{Hybrid SoD}} \\
			\multicolumn{1}{l}{\textbf{Centralised}} & $\bl\bl\bl\bl\sq$ & $\bl\bl\sq\sq\sq$ & $\bl\bl\bl\bl\bl$ & $\bl\bl\bl\bl\sq$ & $\bl\bl\bl\bl\sq$ & $\bl\bl\bl\bl\sq$  \\
			\multicolumn{1}{l}{\textbf{Decentralised}} & $\bl\bl\bl\bl\sq$ & $\bl\bl\bl\sq\sq$  & $\bl\bl\bl\bl\bl$ & $\bl\bl\bl\bl\bl$ & $\bl\bl\bl\bl\bl$ & $\bl\bl\bl\bl\bl$ \\
            \multicolumn{1}{l}{\textbf{Distributed}} & $\bl\bl\bl\bl\sq$ & $\bl\bl\bl\sq\sq$  & $\bl\bl\bl\bl\bl$ & $\bl\bl\bl\bl\bl$ & $\bl\bl\bl\bl\bl$ & $\bl\bl\bl\bl\bl$  \\
 \bottomrule
		\end{tabular}
	}
\end{table}


\section{Conclusion}
\label{sec:Conclusion}
The potential for having an independent and autonomous set of drones, whether we call them FoD or SoD, is gradually becoming necessary, especially with the increased complexities of making real-time decisions by individual or fleets of drones in an overcrowded and regulated airspace.  For such an eventuality, the application of swarm intelligence in the UAVs domain is only a natural progression of the field. In this paper, we forwarded the rationale for drone fleets, the need for them to be independent and autonomous in the wild, and the application of swarm intelligence. We also explained a conceptual architecture to integrate swarm intelligence as a core function, not a merely restricted to a single function like traffic management, it manages and controls a wide range of functions and decisions at the individual drone and fleet level. To support this conceptual architecture, we have also listed different types and collaboration models of SoD along with open issues whose solutions are crucial to the success of the application of swarm intelligence as a core function of the FoDs.

\bibliographystyle{IEEEtran}
\bibliography{IEEEabrv,./main}

\begin{thebibliography}{10}
\providecommand{\url}[1]{#1}
\csname url@samestyle\endcsname
\providecommand{\newblock}{\relax}
\providecommand{\bibinfo}[2]{#2}
\providecommand{\BIBentrySTDinterwordspacing}{\spaceskip=0pt\relax}
\providecommand{\BIBentryALTinterwordstretchfactor}{4}
\providecommand{\BIBentryALTinterwordspacing}{\spaceskip=\fontdimen2\font plus
\BIBentryALTinterwordstretchfactor\fontdimen3\font minus
  \fontdimen4\font\relax}
\providecommand{\BIBforeignlanguage}[2]{{%
\expandafter\ifx\csname l@#1\endcsname\relax
\typeout{** WARNING: IEEEtran.bst: No hyphenation pattern has been}%
\typeout{** loaded for the language `#1'. Using the pattern for}%
\typeout{** the default language instead.}%
\else
\language=\csname l@#1\endcsname
\fi
#2}}
\providecommand{\BIBdecl}{\relax}
\BIBdecl

\bibitem{canis2015unmanned}
B.~Canis, \emph{Unmanned aircraft systems (UAS): Commercial outlook for a new
  industry}.\hskip 1em plus 0.5em minus 0.4em\relax Congressional Research
  Service Washington, 2015.

\bibitem{Beni2005}
\BIBentryALTinterwordspacing
G.~Beni, ``{From Swarm Intelligence to Swarm Robotics},'' in
  \emph{International Workshop on Swarm Robotics, SR 2004}.\hskip 1em plus
  0.5em minus 0.4em\relax Springer, Berlin, Heidelberg, 2005, pp. 1--9.
  [Online]. Available:
  \url{http://link.springer.com/10.1007/978-3-540-30552-1_1}
\BIBentrySTDinterwordspacing

\bibitem{Brambilla2013}
\BIBentryALTinterwordspacing
M.~Brambilla, E.~Ferrante, M.~Birattari, and M.~Dorigo, ``{Swarm robotics: a
  review from the swarm engineering perspective},'' \emph{Swarm Intelligence},
  vol.~7, no.~1, pp. 1--41, jan 2013. [Online]. Available:
  \url{http://link.springer.com/10.1007/s11721-012-0075-2}
\BIBentrySTDinterwordspacing

\bibitem{Barca2013}
\BIBentryALTinterwordspacing
J.~C. Barca and Y.~A. Sekercioglu, ``{Swarm robotics reviewed},''
  \emph{Robotica}, vol.~31, no.~03, pp. 345--359, may 2013. [Online].
  Available: \url{http://www.journals.cambridge.org/abstract_S026357471200032X}
\BIBentrySTDinterwordspacing

\bibitem{Dorigo2013}
\BIBentryALTinterwordspacing
M.~Dorigo, D.~Floreano, L.~M. Gambardella, F.~Mondada, S.~Nolfi, T.~Baaboura,
  M.~Birattari, M.~Bonani, M.~Brambilla, A.~Brutschy, D.~Burnier, A.~Campo,
  A.~L. Christensen, A.~Decugniere, G.~{Di Caro}, F.~Ducatelle, E.~Ferrante,
  A.~Forster, J.~M. Gonzales, J.~Guzzi, V.~Longchamp, S.~Magnenat, N.~Mathews,
  M.~{Montes de Oca}, R.~O'Grady, C.~Pinciroli, G.~Pini, P.~Retornaz,
  J.~Roberts, V.~Sperati, T.~Stirling, A.~Stranieri, T.~Stutzle, V.~Trianni,
  E.~Tuci, A.~E. Turgut, and F.~Vaussard, ``{Swarmanoid: A Novel Concept for
  the Study of Heterogeneous Robotic Swarms},'' \emph{IEEE Robotics {\&}
  Automation Magazine}, vol.~20, no.~4, pp. 60--71, dec 2013. [Online].
  Available: \url{http://ieeexplore.ieee.org/document/6603259/}
\BIBentrySTDinterwordspacing

\bibitem{Urzelai2001}
\BIBentryALTinterwordspacing
J.~Urzelai and D.~Floreano, ``\BIBforeignlanguage{en}{{Evolution of adaptive
  synapses: robots with fast adaptive behavior in new environments.}}''
  \emph{\BIBforeignlanguage{en}{Evolutionary Computation}}, vol.~9, no.~4, pp.
  495--524, jan 2001. [Online]. Available:
  \url{http://www.mitpressjournals.org/doi/abs/10.1162/10636560152642887#.VUnQjeS7zUJ}
\BIBentrySTDinterwordspacing

\bibitem{Bredeche2009}
\BIBentryALTinterwordspacing
N.~Bredeche, E.~Haasdijk, and A.~E. Eiben, ``{On-line, on-board evolution of
  robot controllers},'' in \emph{Proceedings of the 9th international
  conference on Artificial evolution}.\hskip 1em plus 0.5em minus 0.4em\relax
  Strasbourg: Springer-Verlag, oct 2009, pp. 110--121. [Online]. Available:
  \url{http://dl.acm.org/citation.cfm?id=1883723.1883738}
\BIBentrySTDinterwordspacing

\bibitem{Millard2014a}
\BIBentryALTinterwordspacing
A.~G. Millard, J.~Timmis, and A.~F.~T. Winfield, ``{Run-time detection of
  faults in autonomous mobile robots based on the comparison of simulated and
  real robot behaviour},'' in \emph{2014 IEEE/RSJ International Conference on
  Intelligent Robots and Systems}.\hskip 1em plus 0.5em minus 0.4em\relax
  Chicago, IL, USA: IEEE, sep 2014, pp. 3720--3725. [Online]. Available:
  \url{http://ieeexplore.ieee.org/document/6943084/}
\BIBentrySTDinterwordspacing

\bibitem{Saska2014}
\BIBentryALTinterwordspacing
M.~Saska, J.~Chudoba, L.~Precil, J.~Thomas, G.~Loianno, A.~Tresnak, V.~Vonasek,
  and V.~Kumar, ``{Autonomous deployment of swarms of micro-aerial vehicles in
  cooperative surveillance},'' in \emph{2014 International Conference on
  Unmanned Aircraft Systems (ICUAS)}.\hskip 1em plus 0.5em minus 0.4em\relax
  Orlando, FL, USA: IEEE, may 2014, pp. 584--595. [Online]. Available:
  \url{http://ieeexplore.ieee.org/document/6842301/}
\BIBentrySTDinterwordspacing

\bibitem{javaid2012cyber}
A.~Y. Javaid, W.~Sun, V.~K. Devabhaktuni, and M.~Alam, ``Cyber security threat
  analysis and modeling of an unmanned aerial vehicle system,'' in
  \emph{Homeland Security (HST), 2012 IEEE Conference on Technologies
  for}.\hskip 1em plus 0.5em minus 0.4em\relax IEEE, 2012, pp. 585--590.

\bibitem{gupta2016survey}
L.~Gupta, R.~Jain, and G.~Vaszkun, ``Survey of important issues in uav
  communication networks,'' \emph{IEEE Communications Surveys \& Tutorials},
  vol.~18, no.~2, pp. 1123--1152, 2016.

\bibitem{Gregory2013}
\BIBentryALTinterwordspacing
W.~G. Voss, ``Privacy law implications of the use of drones for security and
  justice purposes,'' \emph{International Journal of Liability and Scientific
  Enquiry}, vol.~6, no.~4, pp. 171--192, 2013, pMID: 60848. [Online].
  Available:
  \url{http://www.inderscienceonline.com/doi/abs/10.1504/IJLSE.2013.060848}
\BIBentrySTDinterwordspacing

\bibitem{Elias2012}
\BIBentryALTinterwordspacing
B.~Elias, ``Pilotless drones: Background and considerations for congress
  regarding unmanned aircraft operations in the national airspace system,''
  Congressional Research Service, CRS Report for Congress, September 2012.
  [Online]. Available: \url{http://biotech.law.lsu.edu/crs/R42718.pdf}
\BIBentrySTDinterwordspacing

\bibitem{Chang:2017}
\BIBentryALTinterwordspacing
V.~Chang, P.~Chundury, and M.~Chetty, ``Spiders in the sky: User perceptions of
  drones, privacy, and security,'' in \emph{Proceedings of the 2017 CHI
  Conference on Human Factors in Computing Systems}, ser. CHI '17.\hskip 1em
  plus 0.5em minus 0.4em\relax New York, NY, USA: ACM, 2017, pp. 6765--6776.
  [Online]. Available: \url{http://doi.acm.org/10.1145/3025453.3025632}
\BIBentrySTDinterwordspacing

\bibitem{Lidynia2017}
\BIBentryALTinterwordspacing
C.~Lidynia, R.~Philipsen, and M.~Ziefle, \emph{Droning on About
  Drones---Acceptance of and Perceived Barriers to Drones in Civil Usage
  Contexts}.\hskip 1em plus 0.5em minus 0.4em\relax Cham: Springer
  International Publishing, 2017, pp. 317--329. [Online]. Available:
  \url{https://doi.org/10.1007/978-3-319-41959-6_26}
\BIBentrySTDinterwordspacing

\bibitem{Cavoukian2012}
\BIBentryALTinterwordspacing
A.~Cavoukian, ``Privacy and drones: Unmanned aerial vehicles,'' Information and
  Privacy Commissioner of Ontario, Ontario, Canada, Tech. Rep., August 2012.
  [Online]. Available:
  \url{https://www.publicsafety.gc.ca/lbrr/archives/cnmcs-plcng/cn29822-eng.pdf}
\BIBentrySTDinterwordspacing

\bibitem{amazondrone2016}
\BIBentryALTinterwordspacing
(2016, December) Amazon claims first successful prime air drone delivery.
  [Online]. Available:
  \url{https://www.theguardian.com/technology/2016/dec/14/amazon-claims-first-successful-prime-air-drone-delivery}
\BIBentrySTDinterwordspacing

\bibitem{serge:DBLP:conf/iros/AbdillaRB15}
\BIBentryALTinterwordspacing
A.~Abdilla, A.~Richards, and S.~Burrow, ``Power and endurance modelling of
  battery-powered rotorcraft,'' in \emph{2015 {IEEE/RSJ} International
  Conference on Intelligent Robots and Systems, {IROS} 2015, Hamburg, Germany,
  September 28 - October 2, 2015}.\hskip 1em plus 0.5em minus 0.4em\relax
  {IEEE}, 2015, pp. 675--680. [Online]. Available:
  \url{https://doi.org/10.1109/IROS.2015.7353445}
\BIBentrySTDinterwordspacing

\bibitem{serge:article}
J.~A.~Montenegro, M.~Pinto, and L.~Fuentes, ``An empirical study of the power
  consumption of cryptographic primitives in android,'' 05 2017.

\bibitem{serge:doi:10.1117/1.JRS.10.016030}
\BIBentryALTinterwordspacing
J.~Z. W. C. H. D. H. J. A.~T. Huaibo~Song, Chenghai~Yang, ``Comparison of
  mosaicking techniques for airborne images from consumer-grade cameras,''
  \emph{Journal of Applied Remote Sensing}, vol.~10, pp. 10 -- 10 -- 14, 2016.
  [Online]. Available: \url{http://dx.doi.org/10.1117/1.JRS.10.016030}
\BIBentrySTDinterwordspacing

\bibitem{serge:unknown}
G.~Chmaj and H.~Selvaraj, ``Distributed processing applications for uav/drones:
  A survey,'' 08 2014.

\bibitem{serge:chaumette:hal-01391871}
\BIBentryALTinterwordspacing
S.~Chaumette, ``{Chapter 8: Cooperating UAVs and Swarming},'' in \emph{{UAV
  Networks and Communications}}, S.~C. Kamesh~Namuduri, Jae H.~Kim and J.~P.
  Sterbenz, Eds.\hskip 1em plus 0.5em minus 0.4em\relax {Cambridge University
  Press}, 2016. [Online]. Available:
  \url{https://hal.archives-ouvertes.fr/hal-01391871}
\BIBentrySTDinterwordspacing

\bibitem{wei2013agent}
Y.~Wei, G.~R. Madey, and M.~B. Blake, ``Agent-based simulation for uav swarm
  mission planning and execution,'' in \emph{Proceedings of the Agent-Directed
  Simulation Symposium}.\hskip 1em plus 0.5em minus 0.4em\relax Society for
  Computer Simulation International, 2013, p.~2.

\bibitem{purta2013multi}
R.~Purta, S.~Nagrecha, and G.~Madey, ``Multi-hop communications in a swarm of
  uavs,'' in \emph{Proceedings of the Agent-Directed Simulation
  Symposium}.\hskip 1em plus 0.5em minus 0.4em\relax Society for Computer
  Simulation International, 2013, p.~5.

\bibitem{Madey2014}
\BIBentryALTinterwordspacing
G.~Madey, ``Dynamic predictive simulations of agent swarms (dddas),''
  University of Notre Dame, IN, USA, Defense Technical Information Center
  Report ADA601979, Jan 2014. [Online]. Available:
  \url{http://www.dtic.mil/get-tr-doc/pdf?AD=ADA601979}
\BIBentrySTDinterwordspacing

\bibitem{7303086}
L.~Apvrille, Y.~Roudier, and T.~J. Tanzi, ``Autonomous drones for disasters
  management: Safety and security verifications,'' in \emph{2015 1st URSI
  Atlantic Radio Science Conference (URSI AT-RASC)}, May 2015, pp. 1--2.

\bibitem{Ross2013}
\BIBentryALTinterwordspacing
S.~Ross, N.~Melik-Barkhudarov, K.~S. Shankar, A.~Wendel, D.~Dey, J.~A. Bagnell,
  and M.~Hebert, ``{Learning monocular reactive UAV control in cluttered
  natural environments},'' in \emph{2013 IEEE International Conference on
  Robotics and Automation}.\hskip 1em plus 0.5em minus 0.4em\relax IEEE, may
  2013, pp. 1765--1772. [Online]. Available:
  \url{http://ieeexplore.ieee.org/document/6630809/}
\BIBentrySTDinterwordspacing

\bibitem{Wang2016}
\BIBentryALTinterwordspacing
G.-G. Wang, H.~E. Chu, and S.~Mirjalili, ``{Three-dimensional path planning for
  UCAV using an improved bat algorithm},'' \emph{Aerospace Science and
  Technology}, vol.~49, pp. 231--238, feb 2016. [Online]. Available:
  \url{http://linkinghub.elsevier.com/retrieve/pii/S1270963815003843}
\BIBentrySTDinterwordspacing

\bibitem{Couceiro2013}
\BIBentryALTinterwordspacing
M.~S. Couceiro, D.~Portugal, and R.~P. Rocha, ``{A collective robotic
  architecture in search and rescue scenarios},'' in \emph{Proceedings of the
  28th Annual ACM Symposium on Applied Computing - SAC '13}.\hskip 1em plus
  0.5em minus 0.4em\relax New York, New York, USA: ACM Press, 2013, p.~64.
  [Online]. Available:
  \url{http://dl.acm.org/citation.cfm?doid=2480362.2480377}
\BIBentrySTDinterwordspacing

\bibitem{Pugh2006}
\BIBentryALTinterwordspacing
J.~Pugh and A.~Martinoli, ``{Multi-robot learning with particle swarm
  optimization},'' in \emph{Proceedings of the fifth international joint
  conference on Autonomous agents and multiagent systems - AAMAS '06}.\hskip
  1em plus 0.5em minus 0.4em\relax New York, New York, USA: ACM Press, 2006, p.
  441. [Online]. Available:
  \url{http://portal.acm.org/citation.cfm?doid=1160633.1160715}
\BIBentrySTDinterwordspacing

\bibitem{Vasarhelyi2014}
\BIBentryALTinterwordspacing
G.~Vasarhelyi, C.~Viragh, G.~Somorjai, N.~Tarcai, T.~Szorenyi, T.~Nepusz, and
  T.~Vicsek, ``{Outdoor flocking and formation flight with autonomous aerial
  robots},'' in \emph{2014 IEEE/RSJ International Conference on Intelligent
  Robots and Systems}.\hskip 1em plus 0.5em minus 0.4em\relax IEEE, sep 2014,
  pp. 3866--3873. [Online]. Available:
  \url{http://ieeexplore.ieee.org/document/6943105/}
\BIBentrySTDinterwordspacing

\bibitem{AkramTrustCom2016d}
R.~N. Akram, P.~F. Bonnefoi, S.~Chaumette, K.~Markantonakis, and D.~Sauveron,
  ``Secure autonomous uavs fleets by using new specific embedded secure
  elements,'' in \emph{2016 IEEE Trustcom/BigDataSE/ISPA}, Aug 2016, pp.
  606--614.

\bibitem{Steinmann2016ICNS}
J.~A. Steinmann, R.~F. Babiceanu, and R.~Seker, ``Uas security: Encryption key
  negotiation for partitioned data,'' in \emph{2016 Integrated Communications
  Navigation and Surveillance (ICNS)}, April 2016, pp. 1E4--1--1E4--7.

\bibitem{Blazy2017ICNS}
O.~Blazy, P.-F. Bonnefoi, E.~Conchon, D.~Sauveron, R.~N. Akram,
  K.~Markantonakis, K.~Mayes, and S.~Chaumette, ``An efficient protocol for uas
  security,'' in \emph{2017 Integrated Communications Navigation and
  Surveillance (ICNS)}, 2017.

\bibitem{Won2015AsiaCCS}
\BIBentryALTinterwordspacing
J.~Won, S.-H. Seo, and E.~Bertino, ``A secure communication protocol for drones
  and smart objects,'' in \emph{Proceedings of the 10th ACM Symposium on
  Information, Computer and Communications Security}, ser. ASIA CCS '15.\hskip
  1em plus 0.5em minus 0.4em\relax New York, NY, USA: ACM, 2015, pp. 249--260.
  [Online]. Available: \url{http://doi.acm.org/10.1145/2714576.2714616}
\BIBentrySTDinterwordspacing

\bibitem{mtita2017dasc}
C.~Mtita, M.~Laurent, D.~Sauveron, R.~N. Akram, K.~Markantonakis, and
  S.~Chaumette, ``Serverless protocols for inventory and tracking with a uav,''
  in \emph{2017 IEEE/AIAA 36th Digital Avionics Systems Conference
  (DASC)}.\hskip 1em plus 0.5em minus 0.4em\relax IEEE, 2017.

\bibitem{mtita2016efficient}
C.~Mtita, M.~Laurent, and J.~Delort, ``Efficient serverless radio-frequency
  identification mutual authentication and secure tag search protocols with
  untrusted readers,'' \emph{IET Information Security}, vol.~10, no.~5, pp.
  262--271, 2016.

\bibitem{Akram-ICNS2017a}
R.~N. Akram, K.~Markantonakis, K.~Mayes, P.~Bonnefoi, D.~Sauveron, and
  S.~Chaumette, ``A secure and trusted protocol for enhancing safety of
  on-ground airplanes using uavs,'' in \emph{Integrated Communications
  Navigation and Surveillance}.\hskip 1em plus 0.5em minus 0.4em\relax IEEE,
  2017.

\bibitem{Pigatto2016}
\BIBentryALTinterwordspacing
D.~F. Pigatto, L.~Gon{\c{c}}alves, G.~F. Roberto, J.~F. Rodrigues~Filho, N.~B.
  Floro~da Silva, A.~R. Pinto, and K.~R. Lucas Jaquie Castelo~Branco, ``The
  hamster data communication architecture for unmanned aerial, ground and
  aquatic systems,'' \emph{Journal of Intelligent {\&} Robotic Systems},
  vol.~84, no.~1, pp. 705--723, 2016. [Online]. Available:
  \url{http://dx.doi.org/10.1007/s10846-016-0356-x}
\BIBentrySTDinterwordspacing

\bibitem{Maxa2016DASC}
J.~A. Maxa, M.~S.~B. Mahmoud, and N.~Larrieu, ``Extended verification of secure
  uaanet routing protocol,'' in \emph{2016 IEEE/AIAA 35th Digital Avionics
  Systems Conference (DASC)}, Sept 2016, pp. 1--16.

\bibitem{Akram2017wistp}
R.~N. Akram, K.~Markantonakis, K.~Mayes, P.~F. Bonnefoi, A.~Cherif,
  D.~Sauveron, and S.~Chaumette, ``A secure and trusted channel protocol for
  uavs fleets,'' in \emph{WISTP2017, The 11th WISTP International Conference on
  Information Security Theory and Practice}, Sep 2017.

\bibitem{serge:abebe2017drone}
\BIBentryALTinterwordspacing
E.~Abebe, A.~Beloglazov, D.~Karunamoorthy, J.~Richter, and K.~Steer, ``Drone
  range extension via host vehicles,'' May~23 2017, uS Patent 9,659,502.
  [Online]. Available: \url{https://www.google.com/patents/US9659502}
\BIBentrySTDinterwordspacing

\bibitem{serge:cummings2007operator}
M.~L. Cummings, A.~S. Brzezinski, and J.~D. Lee, ``Operator performance and
  intelligent aiding in unmanned aerial vehicle scheduling,'' \emph{IEEE
  Intelligent Systems}, vol.~22, no.~2, 2007.

\bibitem{serge:DBLP:conf/syscon/BrustAT16}
\BIBentryALTinterwordspacing
M.~R. Brust, M.~I. Akbas, and D.~Turgut, ``{VBCA:} {A} virtual forces
  clustering algorithm for autonomous aerial drone systems,'' in \emph{Annual
  {IEEE} Systems Conference, SysCon 2016, Orlando, FL, USA, April 18-21, 2016},
  2016, pp. 1--6. [Online]. Available:
  \url{https://doi.org/10.1109/SYSCON.2016.7490517}
\BIBentrySTDinterwordspacing

\end{thebibliography}

\end{document}